\documentclass[12pt]{iopart} 
 
\usepackage{graphicx} 
\begin{document} 
 
\title[Elliptic Flow from a Hybrid CGC, Full 3D Hydro and Hadronic Cascade Model]{ 
Elliptic Flow from a Hybrid CGC, Full 3D Hydro and Hadronic Cascade Model 
} 
 
\author{T Hirano$^1$, U Heinz$^2$, D Kharzeev$^3$, R Lacey$^4$, Y Nara$^5$} 
\address{ 
$^1$Department of Physics, the University of Tokyo, Tokyo 113-0033, Japan\\ 
 
$^2$Department of Physics, Ohio State University, 
%191 West Woodruff Avenue, 
Columbus, OH 43210, USA 
 
$^3$Physics Department, Brookhaven  
             National Laboratory, Upton, NY 11973-5000, USA 
 
$^4$Department of Chemistry, SUNY Stony Brook, Stony Brook,  
NY 11794-3400, USA 
 
$^5$Institut f\"ur Theoretische Physik,  
J.\,W.\,Goethe-Universit\"at, Max\,v.\,Laue\,Str.\,1, D-60438 Frankfurt, 
Germany 
} 
 
\ead{hirano@phys.s.u-tokyo.ac.jp} 
 
\begin{abstract} 
We investigate the robustness of the discovery of the perfect fluid   
through comparison of hydrodynamic calculations with the  
elliptic flow coefficient $v_2$ at midrapidity in Au+Au collisions  
at $\sqrt{s_{NN}}=200$ GeV. 
Employing the Glauber model for initial entropy density distributions, 
the centrality dependence of $v_2$ is reasonably reproduced by using 
an ideal fluid description of the early QGP stage 
followed by a hadronic cascade in the late hadronic stage. 
On the other hand, 
initial conditions based on the Colour Glass Condensate model  
are found to generate 
larger elliptic flow due to larger initial eccentricity $\varepsilon$. 
We further predict $v_2/\varepsilon$ at a fixed impact parameter 
as a function of collision energy $\sqrt{s_{NN}}$ up to the LHC energy. 
\end{abstract} 
 
%\pacs{25.75.-q, 25.75.Nq, 12.38.Mh, 12.38.Qk} 
%\maketitle 

One of the important discoveries made at the Relativistic Heavy Ion  
Collider (RHIC) is the large elliptic flow $v_2$ in non-central Au+Au  
collisions \cite{experiments}. At the highest RHIC energy, the observed 
$v_2$ values near midrapidity at low transverse momentum ($p_T$) in 
central and semicentral collisions agree with predictions from ideal 
fluid dynamics \cite{QGP3}. The ideal fluid dynamical description
gradually breaks down, however, as one moves away from midrapidity or 
studies peripheral collisions. This requires a more realistic treatment 
of the early and late stages in dynamical modeling of relativistic 
heavy ion collisions. 
 
We have formulated a dynamical and unified model \cite{HHKLN}, based on 
fully three-dimensional (3D) ideal fluid dynamics \cite{Hirano,HiranoTsuda}, 
toward understanding the bulk and transport properties of the quark gluon 
plasma (QGP). During the fluid dynamical evolution we assume local 
thermal equilibrium. However, this assumption can be expected to hold 
only during the intermediate stage of the collision. In order to extract 
properties of the QGP from experimental data one must therefore supplement
the hydrodynamic description by appropriate models for the beginning and
end of the collision. For the early stage, we employ the Colour Glass 
Condensate (CGC) picture for colliding nuclei and calculate the
produced gluon distributions as input for the initial conditions 
in the fluid dynamical calculation \cite{HN04}. 
%On the other hand,  
During the late stage, local thermal equilibrium is no longer 
maintained due to expansion and dilution of the matter. 
We treat this gradual transition from a locally thermalised system 
to free-streaming hadrons via a dilute interacting hadronic gas by 
employing a hadronic cascade model \cite{jam}.

\begin{figure}[t] 
\begin{flushright} 
\includegraphics[width=0.45\textwidth]{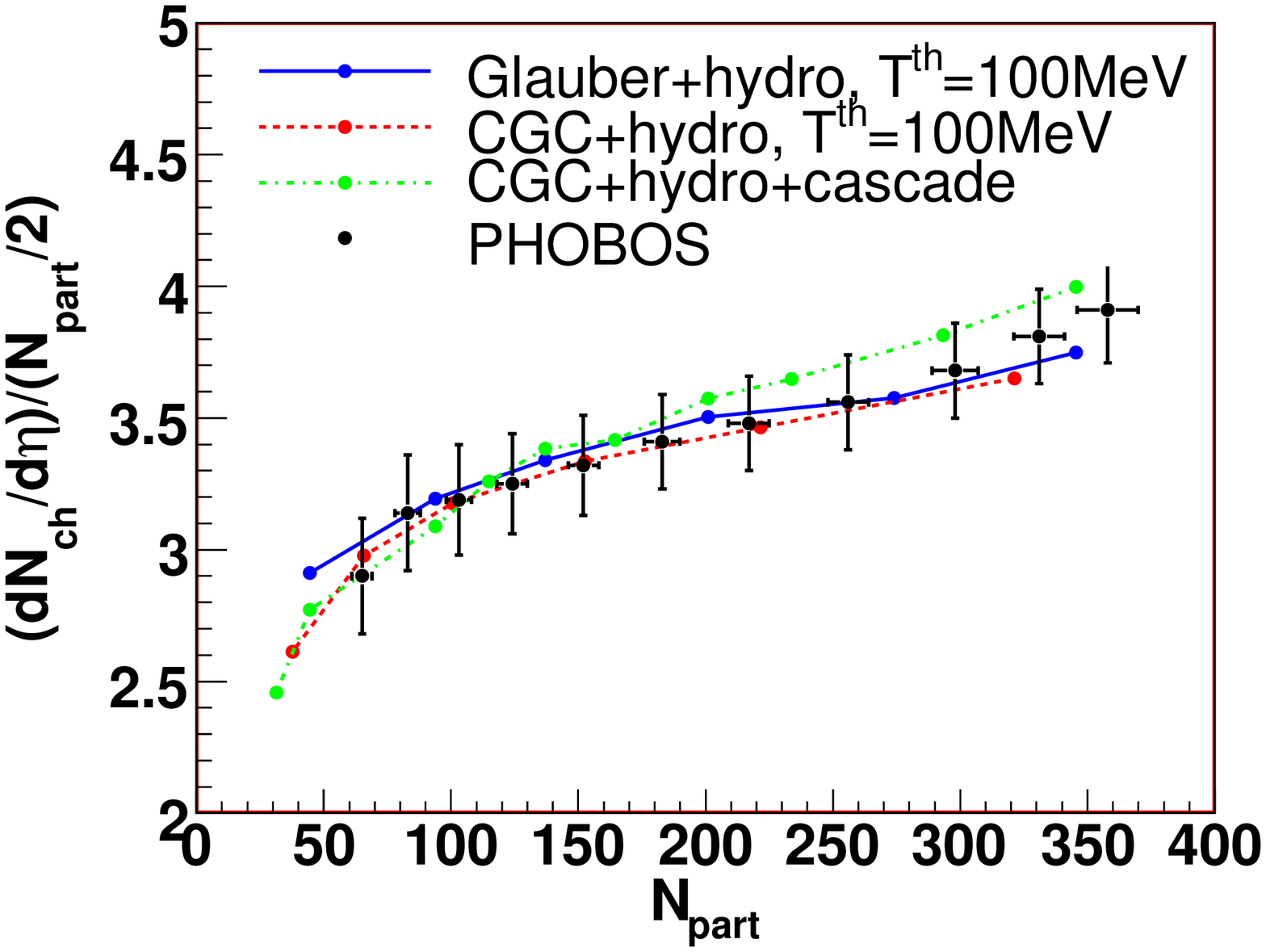} 
\includegraphics[width=0.45\textwidth]{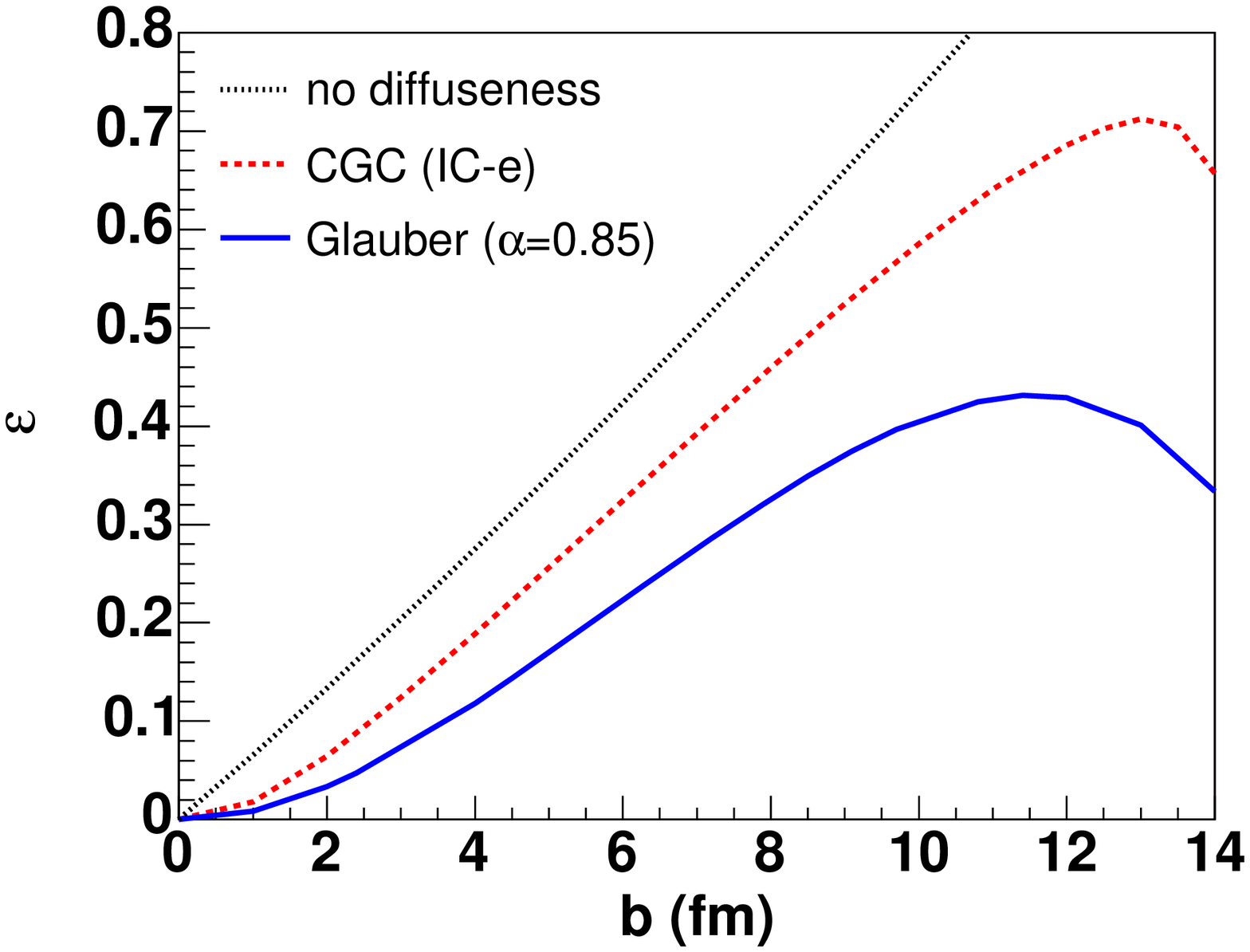} 
\end{flushright} 
\caption{ 
Centrality dependence of charged particle multiplicity per number of 
participant nucleons (left) and impact parameter dependence of  
the eccentricity of the initial energy density distributions (right).  
The solid (dashed) line results from Glauber-type (CGC) initial conditions. 
The dash-dotted line in the left figure results from our hybrid model. 
Experimental data are from PHOBOS \cite{PHOBOS}. The dotted line in the 
right figure assumes a box profile for the initial energy density 
distribution.  
} 
\label{fig:1} 
\end{figure} 
 
We first calculate the centrality dependence of the multiplicity 
to see that the CGC indeed correctly describes the initial entropy 
production and gives proper initial conditions for the fluid dynamical 
calculations. 
For reference we compare with the conventional approach where the 
initial entropy density is parametrized as a superposition of terms 
scaling with the densities of participant nucleons and binary collisions
from the Glauber model. Both CGC and Glauber model initial conditions, 
propagated with 
ideal fluid dynamics, reproduce the observed centrality dependence of 
the multiplicity \cite{PHOBOS}, see Fig.~\ref{fig:1} (left). 
% It should be noted that 
This agreement with the data still holds when the ideal fluid description 
is replaced by a more realistic hadronic cascade below a switching 
temperature  of $T^{\mathrm{sw}}{\,=\,}169$ MeV. In the right panel of 
Fig.~\ref{fig:1} we show the impact parameter dependence of the 
eccentricity of the initial energy density distributions. We neglect 
event-by-event eccentricity fluctuations although these might be 
important for very central and peripheral events \cite{fluctuation}. 
Even though both models correctly describe the centrality dependence 
of the multiplicity, they exhibit a significant difference: The 
eccentricity from the CGC is 20-30\% larger than that from the Glauber 
model \cite{Drescher:2006pi}. The initial eccentricity is thus quite 
sensitive to model assumptions about the initial energy deposition 
which can be discriminated by the observation of elliptic flow. 
 
\begin{figure}[t] 
\begin{flushright} 
\includegraphics[width=0.45\textwidth]{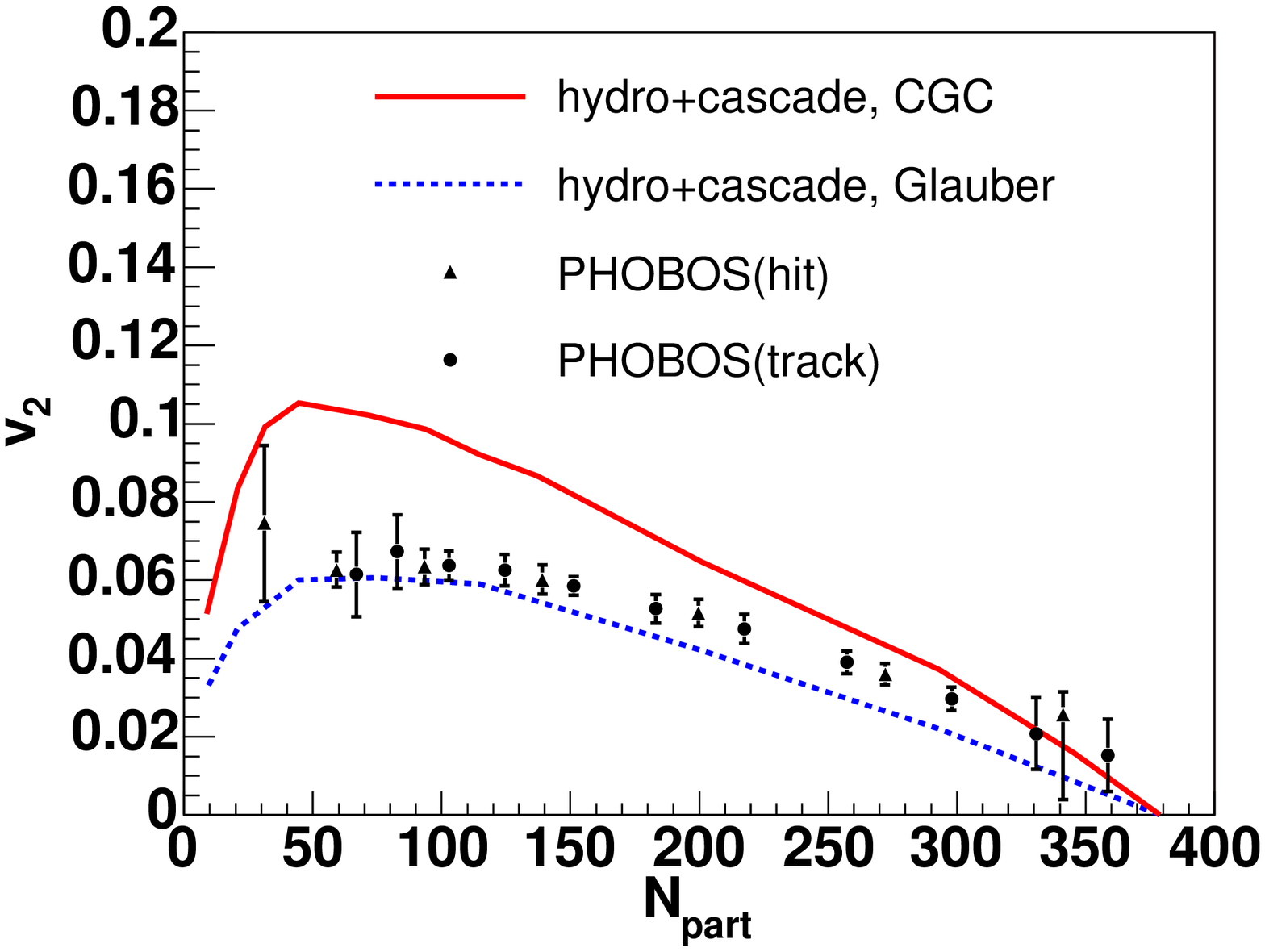} 
\includegraphics[width=0.45\textwidth]{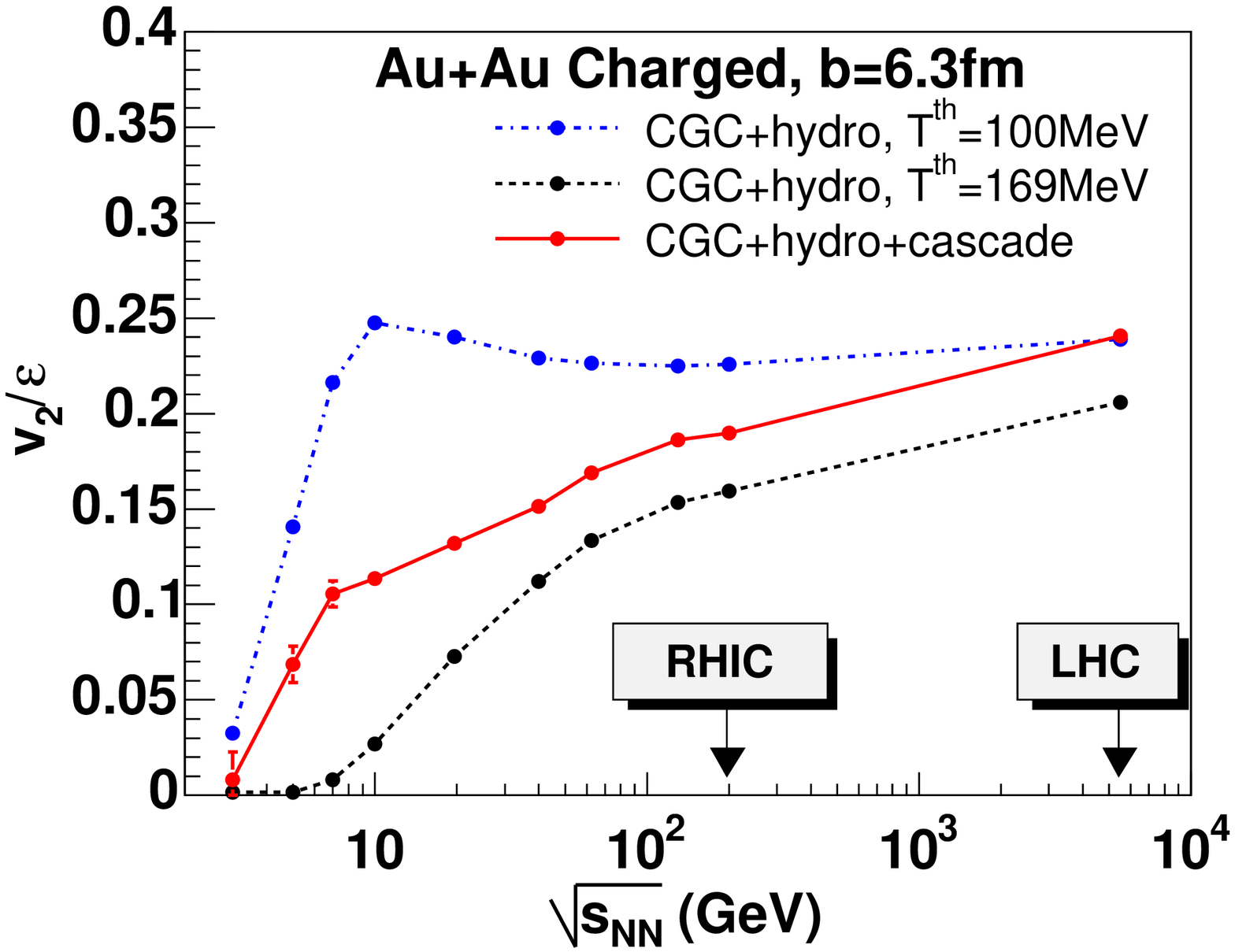} 
\end{flushright} 
\caption{ 
(Left) Centrality dependence of $v_2$. The solid (dashed) line 
results from CGC (Glauber model) initial conditions followed by
ideal fluid QGP dynamics and a dissipative hadronic cascade. The data are 
from the PHOBOS Collaboration \cite{PHOBOS_v2npart}. 
(Right) Excitation function of $v_2/\varepsilon$ in $b=6.3$\,fm Au+Au 
collisions. The solid line results from CGC initial conditions 
followed an ideal QGP fluid and a dissipative hadronic cascade. The 
dashed (dash-dotted) line results from purely ideal fluid dynamics with 
thermal freezeout at $T^{\mathrm{th}}=169$\,MeV ($100$\,MeV). 
} 
\label{fig:2} 
\end{figure}

With Glauber model initial conditions \cite{Kolb}, the predicted $v_2$ 
from ideal fluid dynamics overshoots the peripheral collision data 
\cite{PHOBOS_v2npart}. Hadronic dissipative effects within hadron cascade 
model reduce $v_2$ and, in the Glauber model case, are seen to be sufficient 
to explain the data (Fig.~\ref{fig:2} (left)) \cite{HHKLN}. 
Initial conditions based on the CGC model, however, lead to larger elliptic 
flows which overshoot the data even after hadronic dissipation is accounted
for \cite{HHKLN}, unless one additionally assumes significant 
shear viscosity also during the early QGP stage.
 
Fig.~\ref{fig:2} (right) shows the excitation function of the 
charged particle elliptic flow $v_2$, scaled by the initial 
eccentricity $\varepsilon$, for Au+Au collisions at $b=6.3$\,fm 
impact parameter, using three different models: (i) a pure 3D ideal 
fluid approach with a typical kinetic freezeout temperature
$T^{\mathrm{th}}=100$ MeV where both QGP and hadron gas are treated as
ideal fluids (dash-dotted line); (ii) 3D ideal fluid evolution for the QGP,
with kinetic freezeout at $T^{\mathrm{th}}=169$ MeV and no hadronic 
rescattering (dashed line); and (iii) 3D ideal fluid QGP evolution
followed by hadronic rescattering below $T^\mathrm{sw}=169$\,MeV (solid line). 
%In case (ii) without the hadronic cascade, the spectra 
%are calculated at $T^{\mathrm{th}} = T^{\mathrm{sw}}=169$ MeV. 
% In the conventional fluid dynamical calculations, 
% initial conditions are parametrized 
% so that the final multiplicity is reproduced.  
Although applicability of the CGC model for SPS energies might be 
questioned, we use it here as a systematic tool for obtaining the 
energy dependence of the hydrodynamic initial conditions. By dividing 
out the initial eccentricity $\varepsilon$, we obtain an excitation
function for the scaled elliptic flow $v_2/\varepsilon$ whose shape should 
be insensitive to the fact that CGC initial conditions produce larger
eccentricities and the resulting integrated $v_2$ overshoots the data at 
RHIC. Fig.~\ref{fig:2} shows the well-known bump in $v_2/\varepsilon$ 
at SPS energies ($\sqrt{s_{NN}}\sim10$\,GeV) predicted by the purely 
hydrodynamic approach, as a consequence 
% with $T^{\mathrm{th}}=100$ MeV 
of the softening of the equation of state (EOS) near the quark-hadron 
phase transition region \cite{Kolb2}, and that this structure is completely 
washed out by hadronic 
dissipation \cite{Teaney}, consistent with the experimental data 
\cite{NA49,STAR}. Even at RHIC energy, hadronic dissipation still reduces 
$v_2$ by $\sim$\,20\%. The hybrid model predicts a monotonically increasing 
excitation function for $v_2/\varepsilon$ which keeps growing from RHIC 
to LHC energies \cite{Teaney}, contrary to the ideal fluid approach whose 
excitation function almost saturates above RHIC energies.
 
In summary, we have developed a unified dynamical model, based on fully 
3D ideal QGP fluid dynamics followed by a realistic hadron cascade, to 
describe the space-time evolution of bulk matter in relativistic heavy 
ion collisions. 
% and 
%applied this model  
%to analysis of the current elliptic flow data 
%at RHIC. 
With Glauber initial conditions, hadronic dissipation seems to be able 
to reduce the elliptic flow enough to bring theoretical  
predictions in line with the data at all collision centralities and 
rapidities, leaving little room for additional dissipative effects in 
the QGP. CGC initial conditions, however, yield 
significantly more eccentric sources and produce larger elliptic flow 
than observed, even if late stage hadronic dissipation is accounted for. 
To answer the question whether the QGP is a perfect 
fluid, further systematic studies of the hadron distributions for a 
variety of species, collision centralities, center of mass 
energies and system sizes are needed. Our analysis also points to a 
need for a better understanding of the initial conditions in heavy ion 
collisions if one hopes to use experimental data to constrain the QGP 
viscosity (see recent studies of the source eccentricity from classical 
Yang Mills simulations with a ``universal" saturation scale 
\cite{Lappi:2006xc} and of effects of eccentricity fluctuation 
\cite{fluctuation}). Our approach predicts that the response of 
the system to the initial spatial anisotropy of the produced matter, 
$v_2/\varepsilon$, continues to increase towards LHC energies, 
and that hadronic dissipation washes out the phase-transition 
bump predicted in \cite{Kolb2} and leads to a monotonically excitation 
function for $v_2/\varepsilon$.

%\ack 
This work was supported by the U.S. DOE under contracts  
DE-FG02-01ER41190 (U.H.), DE-AC02-98CH10886 (D.K.) 
and DE-FG02-87ER40331.A008 (R.L.).

\end{document}